\documentclass{article}
\usepackage{ijcai24}

\usepackage{times}
\usepackage{soul}
\usepackage{url}
\usepackage[hidelinks]{hyperref}
\usepackage[utf8]{inputenc}
\usepackage[small]{caption}
\usepackage{graphicx}
\usepackage{amsmath}
\usepackage{amsthm}
\usepackage{booktabs}
\usepackage{algorithm}
\usepackage{algorithmic}
\usepackage[switch]{lineno}

\usepackage{colortbl}
\usepackage{tabstackengine}
\usepackage{multirow}
\usepackage{amssymb}
\usepackage{booktabs}
\usepackage{pifont}
\usepackage{longtable}
\usepackage{array}

\usepackage{tikz}
\usepackage{enumitem}

\title{CompLex: Music Theory Lexicon Constructed by Autonomous Agents for Automatic Music Generation}

\author{
    Zhejing Hu$^1$ \and
    Yan Liu$^1$\thanks{Corresponding author.} \and
    Gong Chen$^1$ \And
    Bruce X.B. Yu$^2$\\
\affiliations 
    $^1$ Department of Computing, The Hong Kong Polytechnic University\\
    $^2$ Zhejiang University-University of Illinois Urbana-Champaign Institute\\
    \emails
    zhejing.hu@connect.polyu.hk, \{yan.liu,gong-cg.chen\}@polyu.edu.hk, xinboyu@intl.zju.edu.cn
}

\begin{document}

\maketitle

\begin{abstract}
Generative artificial intelligence in music has made significant strides, yet it still falls short of the substantial achievements seen in natural language processing, primarily due to the limited availability of music data. Knowledge-informed approaches have been shown to enhance the performance of music generation models, even when only a few pieces of musical knowledge are integrated. This paper seeks to leverage comprehensive music theory in AI-driven music generation tasks, such as algorithmic composition and style transfer, which traditionally require significant manual effort with existing techniques. We introduce a novel automatic music lexicon construction model that generates a lexicon, named CompLex, comprising 37,432 items derived from just 9 manually input category keywords and 5 sentence prompt templates. A new multi-agent algorithm is proposed to automatically detect and mitigate hallucinations. CompLex demonstrates impressive performance improvements across three state-of-the-art text-to-music generation models, encompassing both symbolic and audio-based methods. Furthermore, we evaluate CompLex in terms of completeness, accuracy, non-redundancy, and executability, confirming that it possesses the key characteristics of an effective lexicon.
\end{abstract}

\section{Introduction}

Generative artificial intelligence in music has seen remarkable progress in both academia \cite{copet2024simple,bhandari2025text2midi,hu2025compose} and industry such as Suno and Udio\footnote{www.suno.com; www.udio.com}. These advancements highlight the growing potential of artificial intelligence in the arts and creative industries \cite{wei2023jepoo,karystinaios2023musical,han2024instructme}, offering transformative capabilities for music creation, production, and consumption.

\begin{figure*}[htb]
	\centering
	\includegraphics[width=\linewidth,trim=0 0 0 0,clip]{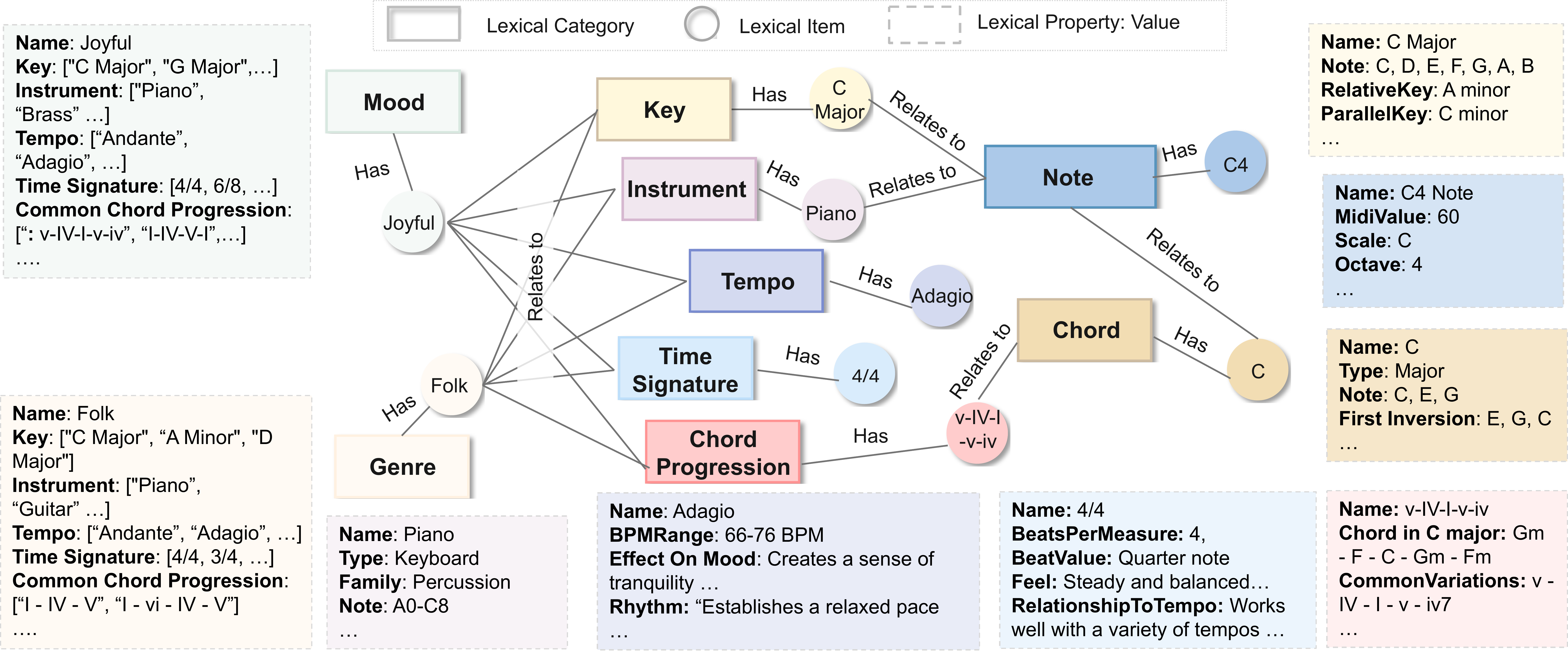}
    \caption{An illustration of CompLex, which contains 9 lexical categories and 37,432 lexical items, each associated with lexical properties and values. The figure displays one item from each category.}
	\label{fig1}
\end{figure*}

Despite significant progress, the development of generative artificial intelligence in music still falls short of the significant achievements seen in natural language processing (NLP) \cite{wei2022chain,farquhar2024detecting}. This gap is mainly due to the limited availability of music data, especially compared to the super large-scale datasets used to train large language models (LLMs) \cite{briot2017deep,ji2023survey}. For example, while LLMs can ensure grammatical accuracy through data-driven pre-training and prompt tuning, large music models (LMMs) face challenges in adhering to music theory using data alone. This discrepancy stems from the vast difference in data availability between language and music models. Billions of new text data points are generated daily, whereas only tens of thousands of new pieces of music, both original and adapted, are created. The limited quantity of music data remains a challenge, as the total number of musical pieces from various periods and styles throughout history amounts to only millions \footnote{https://open.spotify.com/}, starkly contrasting the trillion-level text corpora used to train LLMs \cite{achiam2023gpt}. This limited quantity of music data remains a key bottleneck in the progress of generative music models.

Data augmentation has been proposed as a way to expand music datasets and has achieved significant performance improvements \cite{shorten2019survey}. However, we observe a challenging cycle: effective data augmentation relies on high-performing models, but improving these models depends on high-quality data, which is difficult to generate. Specifically, high-quality music data requires efficient generative models, and if these models perform poorly, the augmented data may introduce noise, undermining further training \cite{shorten2019survey}. Additionally, human auditory perception is highly sensitive to small changes in timing, pitch, and volume \cite{mcdermott2008music}, leading to low tolerance for suboptimal augmented samples. This cycle affects not only music but also many other domains, such as medical imaging \cite{shamshad2023transformers} and finance \cite{cao2022ai}, which also struggle with limited data availability.

In this context, knowledge-informed approaches are essential for enhancing the performance of music generation models. Music theory can significantly improve various aspects of music, including musicality \cite{akama2019controlling}, melody smoothness \cite{wu2020popmnet}, and overall structure \cite{dai2020automatic}. Several components of music theory have already been incorporated into generative models, such as music themes \cite{shih2022theme}, motif-level repetitions \cite{hu2023beauty}, and phrase-level call-and-response structures \cite{hu2024responding}. These works highlight that music theory can indeed boost music performance, even when only a few pieces of musical knowledge are integrated.

In this paper, we aim to leverage comprehensive music theory in AI-driven music generation tasks, such as algorithmic composition and style transfer, which have traditionally demanded considerable manual effort using existing methods. Specifically, we introduce a novel automatic approach to music lexicon construction that generates the music compositional lexicon, CompLex (Figure \ref{fig1}). CompLex is a comprehensive lexicon consisting of 9 categories, 90 properties, and 37,432 lexical items, all derived from just 9 manually input lexical category keywords and 5 prompt templates, significantly reducing reliance on human labor. To facilitate this process, we present LexConstructor, a new multi-agent algorithm designed to automatically detect and mitigate hallucinations during the generation of property-value pairs in the lexicon. CompLex shows impressive performance improvements across three state-of-the-art text-to-music generation models, covering both symbolic and audio-based methods. We also evaluate CompLex in terms of completeness, accuracy, non-redundancy, and executability, confirming that it satisfies the essential characteristics of an effective lexicon.

\begin{figure*}[htbp]
	\centering
	{\centering\includegraphics[width=\linewidth,trim=0 0 0 0,clip]{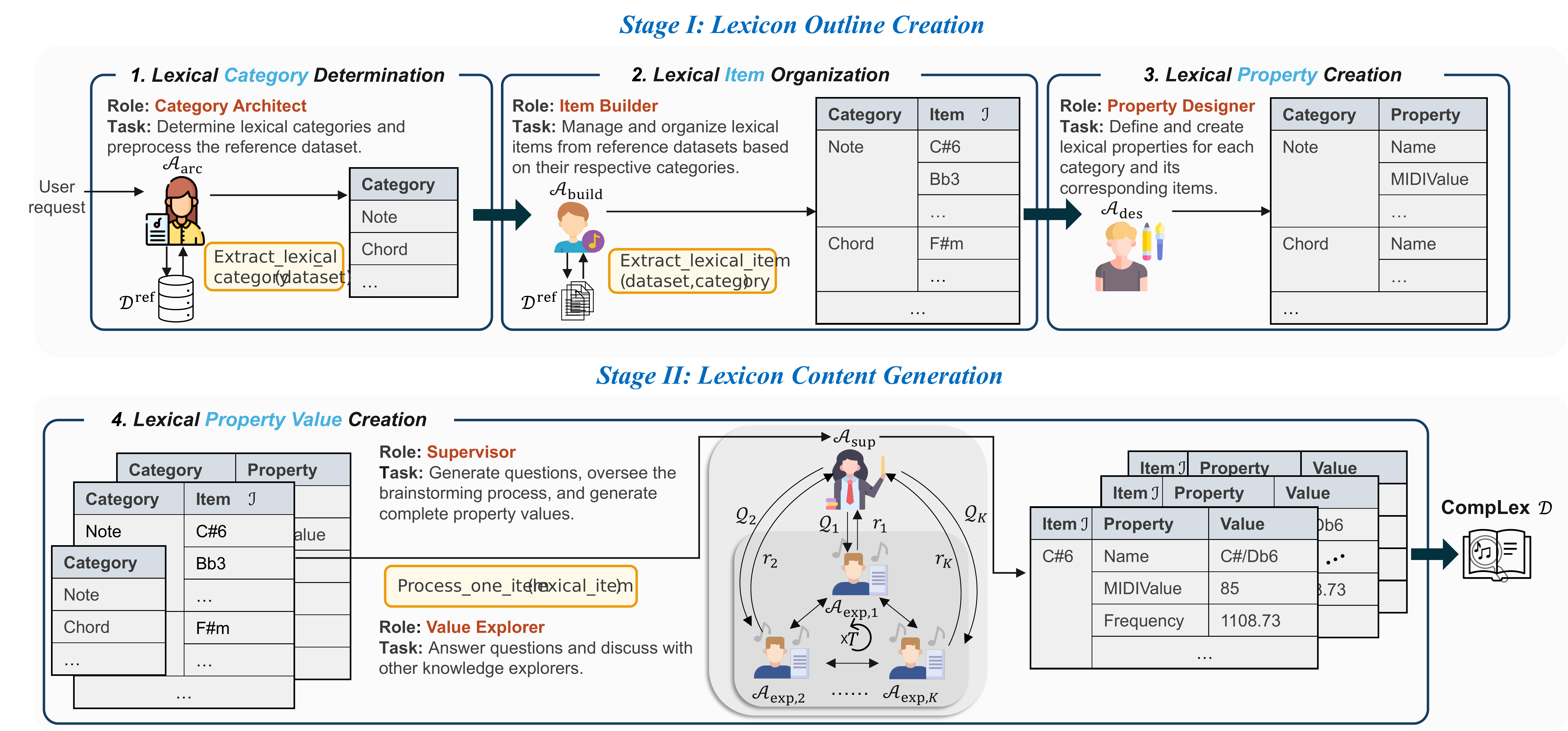}}
	\caption{Overview of the LexConstructor algorithm, which includes two stages: Lexicon Outline Creation and Lexicon Content Generation. Stage I focuses on determining the structure of the lexicon, while Stage II populates it by assigning property-value pairs to items in all categories. The user provides the input request, and then five types of autonomous agents collaborate to create the lexicon.}
\label{framework}
\end{figure*} 

\section{Related Work}
\subsection{Automatic Music Generation}
Automatic music generation has seen significant advancements in recent years, particularly with the development of systems like Suno \cite{suno2024}, which generates high-quality audio-based music from user-provided text input. In academia, models such as MusicGen \cite{copet2024simple} are capable of generating high-quality audio-based music, while Text2Midi \cite{bhandari2025text2midi} specializes in creating symbolic-based music using textual descriptions. Despite these advances, the challenge of limited data in the music domain remains a significant obstacle. To address this, knowledge-informed approaches have proven to be highly beneficial, as they integrate specific aspects of music theory to enhance generation performance, even with limited data samples. Notable works such as Transformer VAE \cite{jiang2020transformer}, Melons \cite{wu2020popmnet}, MusicFrameworks \cite{dai2021controllable}, and MeloForm \cite{lu2022meloform} have demonstrated the impact of incorporating music knowledge into generative models. In this paper, we aim to further advance AI-driven music generation by leveraging comprehensive music theory, specifically through the construction of a detailed music theory lexicon. 

\subsection{Autonomous Agents}
Recently, autonomous agents have garnered significant interest, with notable success achieved by LLMs \cite{wang2024survey,durante2024agent}. The applications of autonomous agents have quickly diversified into various tasks, such as software development \cite{qian2024chatdev}, biomedical discoveries \cite{gao2024empowering} and science debates \cite{cobbe2021training}, and fields such as music \cite{tatar2019musical}, psychology \cite{kovavc2023socialai}, and finance \cite{MayoGW24}. The strengths of the autonomous agent approach—specialization, and robust collaboration—make it particularly suitable for scenarios that require the integration of diverse expertise and coordination of multiple processes. To enable effective collaboration among agents, researchers have developed various communication strategies. For example, a role-reversal strategy is employed to mitigate hallucinations during direct agent communication \cite{qian2024chatdev}. To enhance performance in open-ended discussion scenarios, a debating strategy allows agents to brainstorm and refine ideas through structured debate \cite{chanchateval}. The use of autonomous agents for automatically creating a comprehensive and compact lexicon remains underexplored. Traditional lexicon development relies on manual curation, which is time-consuming. To address this, we propose an automated music lexicon construction algorithm using multiple agents, reducing the need for costly human labor.

\section{Problem Formulation}
We introduce a structured lexicon to organize the vast, unstructured body of music-theory knowledge based on four key components: \textbf{Lexical Category}, \textbf{Lexical Item}, \textbf{Lexical Property}, and \textbf{Lexical Value}.

\paragraph{Lexical Category.}
A lexical category is a high-level concept in music theory that groups related items. Let \(\mathcal{C}\) denote the set of all categories in our framework. Formally, we define:
\begin{equation}
\mathcal{C} = \{\,c_1, c_2, \dots\},
\end{equation}
where each \(c_j \in \mathcal{C}\) represents a distinct music-theory concept. Example categories include ``Chord'', ``Key'', and ``ChordProgression'', among others.

\noindent
\paragraph{Lexical Item.}
A lexical item is a specific instance of a category. For each category \(c \in \mathcal{C}\), let \(\mathcal{I}(c)\) denote the set of items within that category. Formally, we define:
\begin{equation}
\mathcal{I}(c) = \{\,i_1, i_2, \dots\},
\end{equation}
where each \(i_j \in \mathcal{I}(c)\) represents a unique item under the concept \(c\). For instance, under the ``Chord'' category, we might have ``C Major'', ``A Minor'', etc.

\noindent
\paragraph{Lexical Property.}
A lexical property is a named attribute or feature that describes an item within a category. For each category \(c \in \mathcal{C}\), let \(\mathcal{P}(c)\) denote the set of properties associated with items in that category. Formally, we have:
\begin{equation}
\mathcal{P}(c) = \{\,p_1, p_2, \dots\},
\end{equation}
where each \(p_j \in \mathcal{P}(c)\) represents a distinct property under the concept $c$. For example, the ``Chord'' category might include properties such as ``Name'', ``Type'', and ``Quality''.

\noindent
\paragraph{Property-Value Pair.}
Each lexical property \(p_j\) of a lexical item can have a corresponding value, forming a property-value pair. For an item \(i \in \mathcal{I}(c)\) within category \(c\), we define the set of property-value pairs as:
\begin{equation}
\mathcal{V}(i,c) = \{\, (p_1, v_1), (p_2, v_2), \dots \},
\end{equation}
where \(v_j\) is the value associated with the property \(p_j\). For example, for the item ``C Major'' in the ``Chord'' category, we might have:
\[
\mathcal{V}(\text{C Major}, \text{Chord}) = \{\,(\text{Name}, \text{C Major}), (\text{Type}, \text{Triad}), \dots\}.
\]

\noindent
\paragraph{Objective.}
Our objective is to design a model that can automatically identify and extract a comprehensive set of categories \(\mathcal{C}\), the items of each category \(\mathcal{I}(c)\), the properties of each category \(\mathcal{P}(c)\), and generate the corresponding property-value pairs \(\mathcal{V}(i,c)\) correctly from unstructured music-theory knowledge, and finally construct CompLex $\mathcal{D}$ based on a user request $u$.

\section{LexConstructor}
We propose LexConstructor, a multi-agent algorithm composed of several specialized LLM-based agents. Specifically, LexConstructor includes five distinct agent roles, with each agent assigned specific roles and tasks, based on one prompt template. LexConstructor operates in two stages, as shown in Figure \ref{framework}: Lexicon Outline Creation and Lexicon Content Generation. In the Lexicon Outline Creation stage, three types of agents collaboratively develop the lexicon’s structural framework by determining the necessary categories, items, and properties. In the Lexicon Content Generation stage, two types of agents populate the lexicon by assigning property-value pairs to each item within all categories.

\subsection{Stage I: Lexicon Outline Creation}
Inspired by how humans construct a lexicon from scratch, we begin by creating an outline of the lexicon in Stage I. In this stage, LexConstructor focuses on establishing the foundational structure of the lexicon by completing three key subtasks: Lexical Category Determination, Lexical Item Organization, and Lexical Property Creation.

\subsubsection{Lexical Category Determination}

We employ a specialized agent, the \textbf{Category Architect} (\( \mathcal{A}_{\text{arc}} \)), to perform this subtask. The main responsibility of \( \mathcal{A}_{\text{arc}} \) is to determine lexical categories based on user request $u$ and to pre-process the reference MIDI dataset $\mathcal{D}^{\text{ref}}$ into text-based structured information for future subtasks. 
To facilitate this process, \( \mathcal{A}_{\text{arc}} \) utilizes a predefined function \( \texttt{Extract\_lexical\_category}(\mathcal{D}^{\text{ref}}) \), which extracts lexical categories from the MIDI dataset based on the predefined categories. The agent analyzes the dataset, translating the MIDI information into a structured, text-based format, which will be used for the next subtask.

The output of \( \mathcal{A}_{\text{arc}} \) is a set of categories \( \mathcal{C} \), formally defined as:
$
\mathcal{C} = \mathcal{A}_{\text{arc}}(\mathcal{D}^{\text{ref}}, u),
$
where $u$ is the user request and includes 9 category key words in the work, which are mood, genre, key, instrument, tempo, time signature, chord, and note. In addition the reference dataset $\mathcal{D}^{\text{ref}}$ is translated into structured, text-based information.

\subsubsection{Lexical Item Organization}
The \textbf{Item Builder} agent (\( \mathcal{A}_{\text{builder}} \)) is responsible for organizing lexical items within each category identified by \( \mathcal{A}_{\text{arc}} \). Due to the limitations of LLMs in generating large, unique item lists efficiently, \( \mathcal{A}_{\text{builder}} \) leverages the pre-defined function \(\texttt{Extract\_lexical\_item}(c, \mathcal{D}^{\text{ref}})\) to assist in item extraction. Given a reference dataset \( \mathcal{D}^{\text{ref}} \) and a target category \( c \), this function helps the agent identify and extract all relevant items for that category. Formally, this step is represented as: $\mathcal{I}(c) = \mathcal{A}_{\text{build}}(c, \mathcal{D}^{\text{ref}}).$

\subsubsection{Lexical Property Creation}
The \textbf{Property Designer} agent (\( \mathcal{A}_{\text{des}} \)) is tasked with defining and creating lexical properties for each category and its corresponding items. Taking the generated categories and items from earlier steps, \( \mathcal{A}_{\text{des}} \) defines a set of properties \( \mathcal{P}(c) \) for each category. The result is a comprehensive set of properties that describe each category’s items, formally expressed as:
$ \mathcal{P}(c) = \mathcal{A}_{\text{des}}(c, \mathcal{I}(c)).$

\bigskip 

To ensure that the lexicon outline is both compact and comprehensive, the aforementioned subtasks are executed iteratively until all agents determine that no further refinement is needed. This iterative process allows the agents to refine their outputs based on the results from subsequent stages.

\subsection{Stage II: Lexicon Content Generation}
While the outline of the lexicon can be determined through tool usage and by utilizing information directly from the reference dataset, property-value pairs require the exploration of complex knowledge relations that can’t be easily extracted from the dataset or hard-coded into the function. This makes value generation more susceptible to errors, such as inaccurate values assigned to properties. Therefore, we use the LLM's expert knowledge to generate these values during this stage. To further mitigate hallucinations and ensure accurate results, we design a Question-Answering (QA) communication strategy among agents. 

\subsubsection{Lexical Property Value Creation}
We design two different types of agents: the \textbf{Supervisor Agent} (\( \mathcal{A}_{\text{sup}} \)), which oversees the property value generation process, ensures consistency, and manages quality control, and the \textbf{Value Explorer Agents} (\( \mathcal{A}_{\text{exp}} \)), which provide detailed information and populate property values through collaborative exploration. To handle the large number of items to be populated into the lexicon, we define a function \( \texttt{Process\_One\_Item}(\mathcal{I}) \) that allows the agent to process one item at a time, thus avoiding memory constraints of LLMs.

Formally, for each item \( i \in \mathcal{I}(c) \), the value \( \mathcal{V}(i,c) \) is generated as follows:
$\mathcal{V}(i,c) = \mathcal{A}_{\text{sup}} \left( i, \mathcal{P}(c) \right),$
where \( \mathcal{A}_{\text{sup}} \) takes one item \( i \) at a time and generates the value corresponding to the property \( \mathcal{P}(c) \).

The entire lexicon population process involves iterating through all categories and items, which is represented as:

\begin{equation}
\mathcal{D} = \bigcup_{c \in \mathcal{C}} \bigcup_{i \in \mathcal{I}(c)} \mathcal{V}(i,c).
\end{equation}

\subsubsection{Question Answering Communication}
The Supervisor Agent formulates targeted questions, denoted as \( \mathcal{Q} \), and poses them to \( K \) Value Explorer Agents. These agents then collaborate, brainstorming and providing answers to the questions. The Supervisor Agent evaluates the responses and generates the final property-value pair \( \mathcal{V}(i, c) \).

\noindent The process can be formalized as:
\begin{equation}
\left\{
\begin{array}{l}
\mathcal{A}_{\mathrm{sup}} 
\xrightarrow{\{\mathcal{Q}_j\}_{j=1}^K} 
\{\mathcal{A}^{T}_{\text{exp},j}\}_{j=1}^K 
\xrightarrow{\{r_j\}_{j=1}^K} 
\mathcal{A}_{\mathrm{sup}}, \\
\\
\{ \mathcal{A}_{\text{exp},j}^{0} \}_{j=1}^{K} 
\xrightarrow{\{\mathcal{Q}_j\}_{j=1}^{K}} 
\;\cdots\; 
\xrightarrow{\{ r_j^{T-1} \}_{j=1}^{K}} 
\{ \mathcal{A}_{\text{exp},j}^{T} \}_{j=1}^{K}
\end{array}
\right.
\end{equation}

\noindent where:
\begin{itemize}[left=0pt]
    \item \( \mathcal{Q} = \{ q_1, q_2, \ldots \} \) is the set of diverse questions generated by the Supervisor Agent focus on item \( i \).
    \item \( \mathcal{Q}_j \subseteq \mathcal{Q} \) is the subset of questions assigned to \( \mathcal{A}_{\text{exp},j} \).
    \item \( r_j \) is the response from \( \mathcal{A}^{T}_{\text{exp},j} \) to its assigned questions at iteration $T$.
\end{itemize}

\subsection{Text-to-Music Generation Application}  
The lexicon can be applied in text-to-music generation by refining the user prompt. Given a user prompt \( u \), the system queries the lexicon \( \mathcal{D} \) to retrieve related items \( \mathcal{I}(c_u) \), where \( c_u \) is the category corresponding to \( u \). For each item \( i_j \) in \( \mathcal{I}(c_u) \), we extract its properties \( \mathcal{P}(i_j) \), which are then used to find additional related items \( \mathcal{I}(c_k) \) in other categories, expanding the context. The music is generated by using the related lexical items \( \mathcal{I}(c_u) \cup \mathcal{I}(c_k) \). For example, if \( u = \text{"happy mood"} \), items from the ``Mood'' category can be quickly retrieved and then related information such as tempo (120 bpm), key (C major), and chord progression (``I-IV-V-V''), among others, can be further retrieved.

\section{Experiment}
In this section, we provide the experimental setup used to evaluate the performance of CompLex on text-to-music generation tasks and the performance of LexConstructor method.

\subsection{Implementation Details}
\paragraph{Reference Dataset:} We select the \textbf{MidiCaps} dataset \cite{Melechovsky2024} as reference dataset, which is derived from the Lakh MIDI dataset \cite{raffel2016learning}, one of the largest open-source collections of symbolic music data. MidiCaps contains 168,385 MIDI samples.

\paragraph{CompLex Statistics:} The Genre category has 42 items and 13 properties; Mood has 48 items and 9 properties; Key has 24 items and 13 properties; Tempo has 13 items and 10 properties; Time Signature has 86 items and 12 properties; Instrument has 101 items and 8 properties; Chord has 193 items and 8 properties; Chord Progression has 36,797 items and 12 properties; and Note has 128 items and 5 properties.

\paragraph{Language Models:} We test two versions of LLMs: \textbf{GPT-3.5-turbo-0125} and \textbf{GPT-4o}. For lexicon content generation, we utilize \textbf{GPT-4o}, selected for its superior efficiency and larger context window, supporting up to 16,384 tokens.

\paragraph{Model Configurations:} All agents in the system are configured with a temperature setting of 0 \cite{chanchateval}. The agents are assigned specific roles and tasks, with 5 prompt templates where each agent role corresponds to one prompt template, as shown in the Appendix. The user inputs 9 specific lexical categories in the experiment for comparison. The loop in $T$ continues until the agents reach a consensus, at which point no further information is added. For the QA conversation strategy, we utilize $K=3$ value explorers to facilitate brainstorming, enabling a thorough evaluation and refinement of the lexicon entries. For data analysis in the function $\texttt{Extract\_lexical\_category}(\mathcal{D}^{\text{ref}})$, we follow implementations in \cite{Melechovsky2024}.

\subsection{Baselines}

For evaluating \textbf{CompLex} in text-to-music music generation tasks, we assessed its performance using three models: 
\begin{itemize}[left=0pt]
    \item \textbf{Text2MIDI} \cite{bhandari2025text2midi}, a SOTA open-source text-to-symbolic music generation model trained on the MidiCaps dataset;
    \item \textbf{MusicGen} \cite{copet2024simple}, a SOTA open-source text-to-audio music generation model in academia (we used the large version);
    \item \textbf{Suno} \cite{suno2024}, a SOTA black-box text-to-audio music generation model in industry.
\end{itemize}
These baselines represent the most advanced models in both symbolic and audio music generation.

To validate \textbf{LexConstructor}, we compared it against several baselines and state-of-the-art (SOTA) approaches \cite{IslamAP24}, including both single-agent and multi-agent models:
\begin{itemize}[left=0pt]
    \item \textbf{Direct Prompting}, where language models generate the music theory lexicon without explicit guidance.
    \item \textbf{Chain of Thought Prompting (CoT)} \cite{wei2022chain} breaks down tasks into step-by-step subtasks, facilitating the handling of more complex problems.
    \item \textbf{Analogical Reasoning Prompting} \cite{yasunagalarge} directs models to draw upon relevant past experiences to solve similar issues.
    \item \textbf{Multi-Agent Role Reverse} \cite{qian2024chatdev}, where agents reverse roles to collaborate on problem-solving.
    \item \textbf{Multi-Agent Debate} \cite{chanchateval}, where agents engage in debate to find the optimal solution.
\end{itemize}

\subsection{Evaluation Metrics}

\subsubsection{Objective Metrics}
To assess the performance of \textbf{CompLex} on text-to-music generation tasks, we evaluate the following aspects: \textbf{Compression Ratio (CR):} Measures the structure of the music \cite{chuan2018modeling,bhandari2025text2midi}. \textbf{Contrastive Language-Audio Pre-training (CLAP):} Measures how closely the music aligns with the corresponding text caption \cite{bhandari2025text2midi,copet2024simple}. \textbf{Mood Accuracy (MA):} Measures how effectively the generated music reflects the mood indicated in the textual description \cite{Melechovsky2024}. \textbf{Genre Accuracy (GA):} Measures how effectively the generated music reflects the genre in the textual description \cite{Melechovsky2024}.

\begin{table*}[tbp]
  \centering
  \setlength\tabcolsep{14pt}
  \begin{minipage}{\linewidth}
    \scriptsize
    \centering
    \begin{tabular}{ccccccccc}
    \toprule
    \multirow{2}{*}{\textbf{Baseline}} & \multirow{2}{*}{\textbf{Method}} & \multicolumn{4}{c}{\textbf{Objective Metrics}} & \multicolumn{2}{c}{\textbf{Subjective Metrics}} \\
    \cmidrule(lr){3-6} \cmidrule(lr){7-8}
    & & \textbf{CR}$\uparrow$ & \textbf{CLAP}$\uparrow$ & \textbf{MA}$\uparrow$ & \textbf{GA}$\uparrow$ & \textbf{REL}$\uparrow$ & \textbf{OVL}$\uparrow$ \\
    \midrule
    \multirow{3}{*}{\shortstack{Text2MIDI \\ \cite{bhandari2025text2midi}}} 
        & W/o Enhancement & 2.02 & 0.17 & 0.37 & 0.66 & 70.13$\pm$\scriptsize 7.22 & 73.54$\pm$\scriptsize 5.14 \\
        & LLM-Enhanced & 2.07 & 0.24 & 0.41 & 0.66 & 76.54$\pm$\scriptsize 4.55 & 75.50$\pm$\scriptsize 2.93 \\
        & \textbf{CompLex-Enhanced} & \textbf{2.14} & \textbf{0.35} & \textbf{0.47} & \textbf{0.67} & \textbf{78.10}$\pm$\scriptsize \textbf{4.37} & \textbf{77.46}$\pm$\scriptsize \textbf{4.71} \\
    \midrule
    \multirow{3}{*}{\shortstack{MusicGen \\ \cite{copet2024simple}}} 
        & W/o Enhancement & -- & 0.22 & 0.19 & 0.46 & 75.83$\pm$\scriptsize 4.43 & 78.17$\pm$\scriptsize 3.57 \\
        & LLM-Enhanced & -- & 0.25 & 0.26 & 0.47 & 80.17$\pm$\scriptsize 7.72 & 81.20$\pm$\scriptsize 7.96 \\
        & \textbf{CompLex-Enhanced} & -- & \textbf{0.33} & \textbf{0.28} & \textbf{0.51} & \textbf{82.67}$\pm$\scriptsize \textbf{4.29} & \textbf{82.17}$\pm$\scriptsize \textbf{4.45} \\
    \midrule
    \multirow{3}{*}{\shortstack{Suno \\ \cite{suno2024}}}
        & W/o Enhancement & -- & 0.39 & 0.44 & 0.67 & 82.79$\pm$\scriptsize 3.79 & 86.75$\pm$\scriptsize 4.63 \\
        & LLM-Enhanced & -- & 0.39 & 0.43 & 0.52 & 83.88$\pm$\scriptsize 6.05 & 88.67$\pm$\scriptsize 3.93 \\
        & \textbf{CompLex-Enhanced} & -- & \textbf{0.46} & \textbf{0.57} & \textbf{0.79} & \textbf{91.38}$\pm$\scriptsize \textbf{3.84} & \textbf{93.58}$\pm$\scriptsize \textbf{4.13} \\
    \bottomrule
  \end{tabular}
  \caption{Performance of CompLex and other methods across different text-to-music generation models. CompLex-Enhanced method consistently outperforms other methods, indicating the effectiveness of CompLex. The Compression Ratio (CR) metric can only be applied to symbolic music representations (e.g., MIDI) and is not suitable for audio-based music.}
    \label{tablelex}
  \end{minipage}
\end{table*}

To assess the performance of \textbf{LexConstructor} and other baselines, we evaluate the effectiveness of the generated lexicon based on the following aspects: \textbf{Completeness:} Measures how thoroughly the generated lexical items in each category align with the actual items in music theory. The completeness score is averaged across all categories, with higher scores indicating a greater coverage of the relevant items. \textbf{Accuracy:} Measures the alignment between the generated MIDI pitch values and frequency values relative to their corresponding note names. Higher scores reflect better accuracy and adherence to established music theory. \textbf{Non-Redundancy:} Measures the level of non-duplication in the generated lexical items, calculated as the ratio of unique items to the total number of items. Higher scores indicate fewer redundant entries. \textbf{Executability:} Measures the lexicon’s ability to be loaded error-free. Higher scores indicate greater reliability and functionality.

\subsubsection{Subjective Metrics}

Human evaluators assess the generated music from the music generation model based on \textbf{Relevance to Text Input (REL)} and \textbf{Overall Quality (OVL)} \cite{kreuk2022audiogen,copet2024simple} from 1 to 100 where we emphasize musicality in OVL.

An online survey was distributed via social media, receiving 24 valid responses after excluding invalid ones (e.g., participants selecting the same option for all questions or submitting incomplete responses). All participants were fully informed about the purpose of the study and consented to the use of their data for research purposes. Detailed information on the design of the subjective experiment is provided in the Appendix.

\section{Results and Discussion}

This section presents the results and discussion of our experiments. First, we demonstrate the impact of CompLex on text-to-music generation tasks by measuring the quality of the output music. Next, we demonstrate the effectiveness of LexConstructor by evaluating the quality of the generated composition lexicon in terms of completeness, accuracy, non-redundancy, and executability. Additionally, we conduct an ablation analysis to validate the effectiveness of the multi-agent design to reduce redundancy and the QA communication to mitigate hallucinations.

\subsection{Effectiveness of CompLex}
We evaluate the performance of CompLex by refining the user text input on text-to-music generation models. We compare three different knowledge enhancement methods: \textbf{W/o Enhancement}, \textbf{LLM-Enhanced}, and \textbf{CompLex-Enhanced}, which use raw input without any refinement, LLMs, and CompLex to refine the user input. In these methods, the user specifies mood or genre requests.

Table~\ref{tablelex} summarizes the performance of the three baseline models across these methods. The metrics include Compression Ratio (measuring structural compactness), CLAP (measuring alignment between text and audio), Mood Accuracy, Genre Accuracy, Relevance to Input, and Overall Musicality. Across all baseline models, the CompLex-Enhanced method consistently achieves higher performance in all metrics, highlighting the benefits of integrating a music theory lexicon for generating high-quality music. These improvements are primarily driven by the lexicon’s ability to provide specific and structured guidance during the generation process.

The LLM-Enhanced method generally underperforms compared to the CompLex-Enhanced method, primarily due to the lack of domain-specific music theory integration. Additionally, Suno's performance shows improvements when the lexicon is incorporated, underscoring the importance of leveraging structured music knowledge to achieve improvement in both objective and subjective metrics. These findings demonstrate the critical role of CompLex in advancing AI-driven music generation.

\subsection{Effectiveness of LexConstructor}

\begin{table*}[tbp]
  \centering
  \setlength\tabcolsep{12pt}
  \begin{minipage}{\linewidth}
    \scriptsize
    \centering
    \begin{tabular}{llccccccc}
    \toprule
    \textbf{Backbone} & \textbf{Method} &\textbf{Completeness}$\uparrow$ &  \textbf{Accuracy}$\uparrow$ &\textbf{Non-Redundancy}$\uparrow$&\textbf{Executability}$\uparrow$\\
    \hline
    \multirow{8}{*}{\shortstack{GPT-3.5\\-Turbo}} &\textit{Single-Agent:}&&&&\\
    & \ \ \ \ Direct Prompting& 0.54&0.22& 0.05 & 0.53\\
                              &\ \ \ \ CoT Prompting \cite{wei2022chain}& 0.59&0.35& 0.11 & 0.58\\
                          &\ \ \ \ Analogical Prompting \cite{yasunagalarge}& 0.59&0.38& 0.13 & 0.60\\
                          &\textit{Multi-Agent:}&&&&\\
                              &\ \ \ \ Role Reverse \cite{qian2024chatdev} &0.65&0.64& 0.19 & 0.66\\
                               &\ \ \ \ Debate \cite{chanchateval} &0.72&0.68& 0.21 & 0.68\\   
                          &\ \ \ \  \textbf{LexConstructor (Ours)} &  \textbf{0.80}&\textbf{0.78}&\textbf{0.84} & \textbf{0.75}\\
    \hline
    \multirow{8}{*}{GPT-4o} &\textit{Single-Agent:}&&&&\\
    & \ \ \ \ Direct Prompting &0.60&0.32& 0.21 & 0.55\\
& \ \ \ \ CoT  Prompting \cite{wei2022chain}&0.64&0.47& 0.24 & 0.64\\
& \ \ \ \ Analogical Prompting \cite{yasunagalarge}& 0.65&0.48& 0.28 & 0.63\\
&\textit{Multi-Agent:}&&&&\\
& \ \ \ \ Role Reverse \cite{qian2024chatdev}& 0.72&0.72& 0.31 & 0.72\\
& \ \ \ \ Debate \cite{chanchateval} & 0.76&0.75& 0.30 & 0.74\\
& \ \ \ \ \textbf{LexConstructor (Ours)} &\textbf{0.83}&\textbf{0.88}&\textbf{0.95} & \textbf{0.82}\\
  \bottomrule
\end{tabular}
  \caption{Performance of LexConstructor with baseline models, showing metrics completeness, accuracy, non-redundancy, and executability. LexConstructor consistently outperforms baseline models across all metrics, demonstrating its effectiveness in generating a high-quality music theory lexicon.}
  \label{table2}
  \end{minipage}
\end{table*}

Table~\ref{table2} compares the performance of LexConstructor with other baseline models across two different backbones. The evaluation metrics include completeness, accuracy, non-redundancy, and executability. LexConstructor outperforms all baseline models in every metric, highlighting its superior efficiency in generating a music theory lexicon. In addition, the improvements demonstrate LexConstructor's ability to mitigate hallucinations. These gains are largely due to the multi-agent role-playing system, where each agent specializes in specific tasks, and the Question-Answering (QA) communication strategy, which effectively mitigates hallucinations during value generation. The multi-agent approach consistently outperforms single-agent models, underscoring the effectiveness of using multiple agents for complex tasks. Overall, LexConstructor shows strong potential for automatically constructing high-quality lexicons, with applications extending to other domains in the future.

\subsection{Ablation Analysis}

\begin{figure}[htb]
	\centering
	{\centering\includegraphics[width=\linewidth,trim=0 0 0 0,clip]{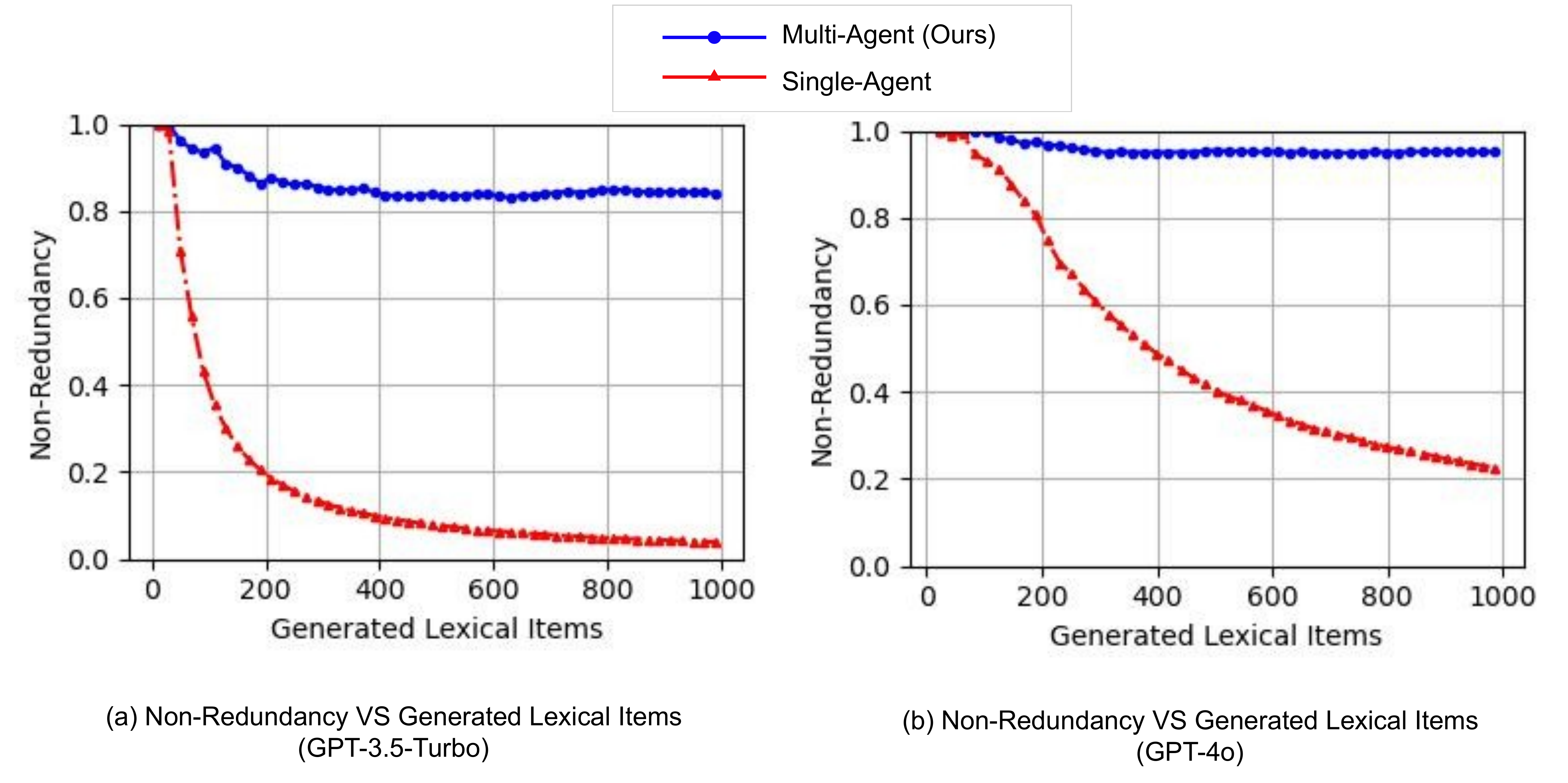}}
	\caption{Ablation Analysis of the Multi-Agent Model: Our multi-agent approach generates lexical items with a higher non-redundancy rate, indicating a better performance.}	
	\label{qua1}
\end{figure}

\paragraph{Ablation Analysis of Multi-Agent:} Figure \ref{qua1} illustrates our model’s efficiency in terms of non-redundancy, showing its performance while generating 1000 lexical items with property values. Single-agent models begin to produce redundant outputs after generating about 50 unique items, with the issue being even more pronounced in the GPT-3.5-Turbo model. The superior performance of our approach can be attributed to its unique design, which incorporates specialized tool usage and task-specific subtasks. By focusing on generating one lexical item at a time, our method avoids referencing previously produced content. While a few duplicates still occur, this is due to the probabilistic nature of GPT models, which sometimes generate content similar to prior outputs.

\paragraph{Ablation Analysis of QA Communication Strategy:} Table \ref{4} shows different communication strategies among agents. It indicates that the QA strategy outperforms other communication strategies, achieving the best performance in mitigating hallucinations that occur during property-value creation. This is because the diverse questions cover different aspects of the concept, and the value explorers collaboratively exchange and refine their answers until reaching a consensus, leading to more accurate and coherent results.

\begin{table}[tbp]
  \centering
  \setlength\tabcolsep{8pt}
  \begin{minipage}{\linewidth}
    \scriptsize
    \centering
    \begin{tabular}{lcc}
        \toprule

            \multirow{2}{*}{\textbf{Strategy}} &  \multicolumn{2}{c}{\textbf{Accuracy}}\\
    \cmidrule(lr){2-3}
    & \textbf{Pitch Value}$\uparrow$ & \textbf{Pitch Frequency}$\uparrow$\\
        \midrule
        w/o Communication& 0.28 & 0.34\\
        w/ Role Reverse& 0.72 & 0.72\\
        w/ Debating& 0.75 & 0.74 \\
        \textbf{w/ QA}& \textbf{0.85} & \textbf{0.91} \\
        \bottomrule
    \end{tabular}
    \caption{Ablation Analysis of the QA Communication Strategy: It generates property values with higher accuracy, effectively mitigating hallucinations.}
    \label{4}
  \end{minipage}
\end{table}

\subsection{Case Study on Hallucination Mitigation}
\begin{figure}[htbp]
	\centering
	{\centering\includegraphics[width=\linewidth,trim=0 0 0 0,clip]{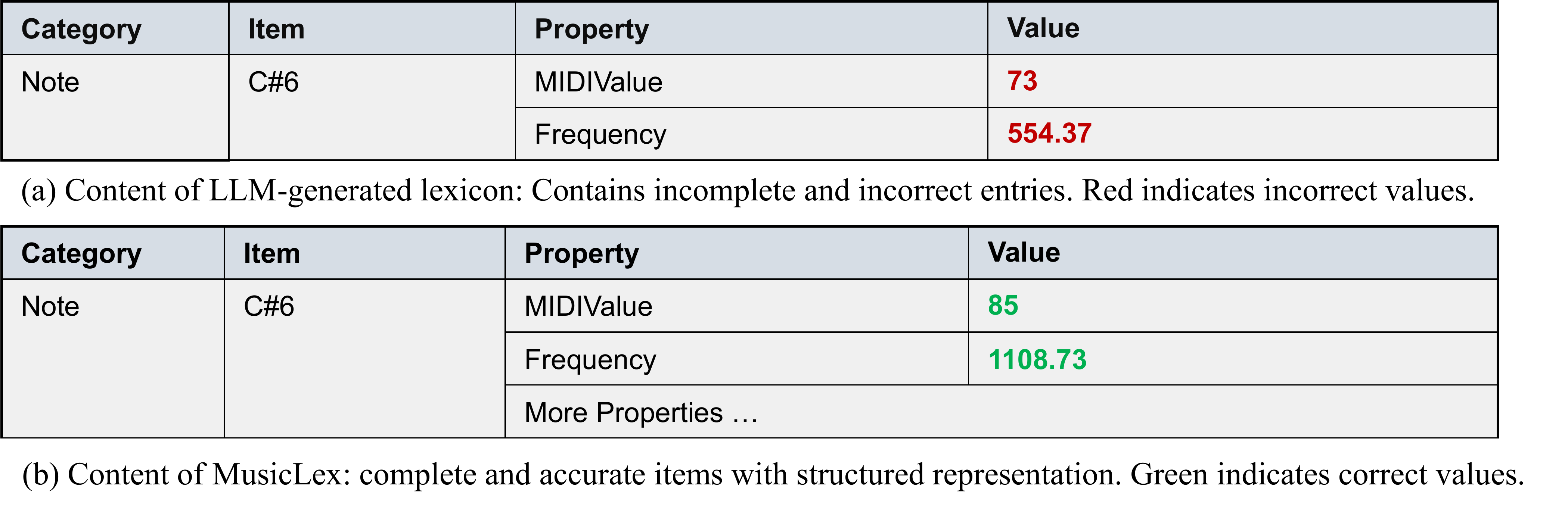}}
	\caption{A case study demonstrating LexConstructor mitigate hallucinations.}
\label{case}
\end{figure}

Figure~\ref{case} presents a case study illustrating how LexConstructor effectively mitigates harmonic hallucinations. The note information in CompLex is correct and complete while generated from LLM-generated lexicon is incorrect and incomplete.

\section{Conclusion}
In this work, we introduce a novel automatic music lexicon construction model that generates CompLex, a comprehensive lexicon containing 9 lexical categories, 90 lexical properties, and 37,432 lexical items, all derived from just 9 manually input category keywords and 5 prompt templates. We also propose a new multi-agent algorithm, LexConstructor, that generates the lexicon and automatically detects and mitigates hallucinations during the generation of property-value pairs for each lexical item. CompLex demonstrates significant performance improvements across three state-of-the-art text-to-music generation models, encompassing both symbolic and audio-based methods. Furthermore, we evaluate CompLex in terms of completeness, accuracy, non-redundancy, and executability, confirming that it meets the key characteristics of an effective lexicon. CompLex can be applied to a broader range of tasks within the music domain, including algorithmic composition, and style transfer. Currently, CompLex focuses on the structure of music composition, in the future, we plan to expand its scope to encompass expressive elements of music performance.

\newpage
\section*{Acknowledgements}
This work is supported by the project Towards Next-generation Artificial Auditory System with Brain-inspired Technologies (C5052-23GF).
\bibliographystyle{named}
\bibliography{ijcai24}

\end{document}